\documentclass[a4paper,12pt]{article}

\usepackage{amsmath,amsfonts,amssymb}
\usepackage{cite}

\usepackage{graphicx}

\begin{document}

\renewcommand*{\thefootnote}{\fnsymbol{footnote}}

\begin{center}
{\Large\bf The Born rule for quantum probabilities from Newton's third law}
\end{center}
%\bigskip
%\vspace{-0.1cm}
\begin{center}
{S.S. Afonin\footnote{E-mail: \texttt{s.afonin@spbu.ru}}
}
\end{center}

\renewcommand*{\thefootnote}{\arabic{footnote}}
\setcounter{footnote}{0}

\begin{center}
  Saint Petersburg State University, 7/9 Universitetskaya nab.,
  St.Petersburg, 199034, Russia\\
  \vspace*{0.15cm}
  and\\
  \vspace*{0.15cm}
  National Research Center "Kurchatov Institute": Petersburg Nuclear Physics Institute,
  mkr. Orlova roshcha 1, Gatchina, 188300, Russia
\end{center}

%\bigskip

\begin{abstract}
According to the Born rule, the probability density in quantum theory is determined by the square of the wave function.
A generally accepted derivation of this rule has not yet been proposed. In the given work, a simple physical picture is constructed
within which the Born rule arises in a natural way. In the proposed scheme, the interaction of a particle with a measuring apparatus
is equivalent to creation of a "mirror image" of particle wave function in the space region of interaction. The observable quantity
is the product of the particle wave function and its "image". The phase of the latter is reversed due to
Newton's third law, thus leading to the Born rule.
\end{abstract}

\bigskip

\section{Introduction}

Recently the fascinating experiments carried out by M.-C. Chen et al.~\cite{chen}
and Z.-D. Li et al.~\cite{li} (following the theoretical scheme designed by M.-O. Renou  et al.~\cite{renou}) have clearly demonstrated that the complex numbers are not
just a convenient tool for description of quantum phenomena, they are fundamental for Quantum Mechanics (QM), at least within the Copenhagen interpretation (see,
e.g., a brief popular exposition of this issue in~\cite{renou2,avella}). These experimental results suggest to reflect once more on the basic fundamental
laws of quantum theory. Although a countless literature on this topic has appeared over the last century, a rich variety of viewpoints and interpretations
creates the feeling that that the presentation of the basic physical concepts of QM in numerous textbooks in this field is still missing something very important.

In the given note, a physical conjecture for this possible missing point is put forward. The discussion will be at a basic level, so that it
can be understood by a broad audience.

Practically all quantitative predictions of quantum physics are based on the Born rule, $P=|\Psi|^2$, first formulated by Max Born in 1926~\cite{born}.
In the standard Copenhagen interpretation of QM and in most alternative ones, this rule (or some version of it) is just postulated. There were
many attempts to derive the Born rule from some other postulates of quantum theory (perhaps the most known attempts were made
in~\cite{gleason,hartle,farhi,deutsch,saunders,zurek}, see also the relevant discussions and references in~\cite{landsman,schlosshauer,lawrence} and recent proposals in~\cite{hossenfelder,gogioso}), however, no generally accepted derivation has been proposed so far. For almost a century, the Born rule remains a basic postulate in standard textbooks on quantum theory.

Most approaches to derivation of Born's rule are based on the belief that this rule is encoded in the mathematical structure of QM (the Gleason theorem~\cite{gleason}
is a canonical example of this kind of thinking) and/or make use of some additional hypotheses (especially in alternative formulations of QM) which are not amenable
to experimental verification. However, it should be understood that the mathematical structure of QM based on complex Hilbert spaces represents only a formal mathematical
model for description of the underlying quantum physics. I believe that the Born rule must have a clear {\it physical} origin related to measurement process and independent
of any particular interpretation of QM or additional hypotheses lying beyond the experimental scope.
The purpose of the present note is to offer a physical explanation for the emergence of Born's rule in quantum theory.

\section{The Born rule in quantum mechanics}

Let us begin with a very brief reminder of how the Born rule arises in the standard QM.
If the linear operator $\hat{L}$ describes a measurement of physical quantity $\lambda$ in the state $\Psi(y)$, the fundamental QM relation connecting
them is
\begin{equation}
\label{1}
\hat{L}\Psi=\lambda\Psi.
\end{equation}
Decomposing $\Psi$ in the orthogonal eigenfunctions $\Psi_k$ of self-conjugate operator $\hat{L}$ (for simplicity, only the discrete spectrum is considered),
one has
\begin{equation}
\label{2}
\hat{L}\Psi=\hat{L}\sum_k c_k\Psi_k=\sum_k \lambda_k c_k\Psi_k.
\end{equation}
Substituting~\eqref{2} into~\eqref{1}, multiplying the result by $\Psi^*=\sum_k c_k^*\Psi_k^*$, and integrating over $y$, one obtains the mathematical expectation
for $\lambda$,
\begin{equation}
\label{3}
\bar{\lambda}=\int\!\Psi^*\hat{L}\Psi\, dy=\sum_k |c_k|^2\lambda_k,
\end{equation}
where the standard normalization, $\int\!\Psi^*\Psi\, dy=1$, is used. Thus, by the classical definition of probability, the probability for $\lambda$
to take the value $\lambda_k$ is
\begin{equation}
\label{4}
P_k=|c_k|^2.
\end{equation}
In the case of continuum spectrum, we would obtain the probability density $\rho(\lambda)$: $\rho(\lambda)d\lambda=|c(\lambda)|^2 d\lambda$. The coefficients
$c_k$ and $c(\lambda)$ represent the wave functions describing the state of a particle (or some composite quantum system) in terms of variables $\lambda$.

The physical interpretation of the relation~\eqref{1} is well known: The linear operator $\hat{L}$ describes the action of a measuring apparatus on a
prepared state $\Psi$ and the observable $\lambda$ is a number which the classical "pointer" of this apparatus "shows" after the measurement. Regarding  the
physical interpretation of the relation~\eqref{3}, the given relation\footnote{Written in the form
$\langle\hat{L}\rangle=\text{tr}(\hat{L}\rho)$, this relation is one of the definitions of Born's rule.} is usually considered as a mathematical
consequence of the statement~\eqref{1}. Below it is argued that the basic fundamental QM statement is expressed by the relation~\eqref{3}
because it fully expresses the measurement process, while~\eqref{1} represents a mathematical consequence of~\eqref{3}, a consequence with a "truncated" physical
information on quantum measurement. More specifically, the operation of convolution of~\eqref{2} with $\Psi^*$ has the physical meaning of taking into account
the "response" of measuring apparatus. Within the emerging interpretation, the relation~\eqref{3} leads to a clear
physical understanding of the origin of Born's rule~\eqref{4}.

\section{Newton's third law in quantum measurement}

The underlying idea will be demonstrated with the help of a simple qualitative example. Consider a free particle described by the wave function
\begin{equation}
\label{5}
\Psi_\text{part}(p,x)=f_\text{part}(p,x)e^{i\varphi},
\end{equation}
where $\varphi$ is the phase and $f_\text{part}(p,x)$ determines a shape of corresponding wave packet in the momentum and coordinate space
(we do not fix a definite basis in the corresponding Hilbert space).
For ease of visualization, let us depict the interaction of this  particle with a measuring apparatus as a contact interaction with a "wall",
see Fig.~1. In reality, this "wall" represents a part of detector and the ingoing particle interacts with some atom or elementary particle inside this detector.

\begin{figure}[!ht]
\begin{center}
    \includegraphics[width=\textwidth]{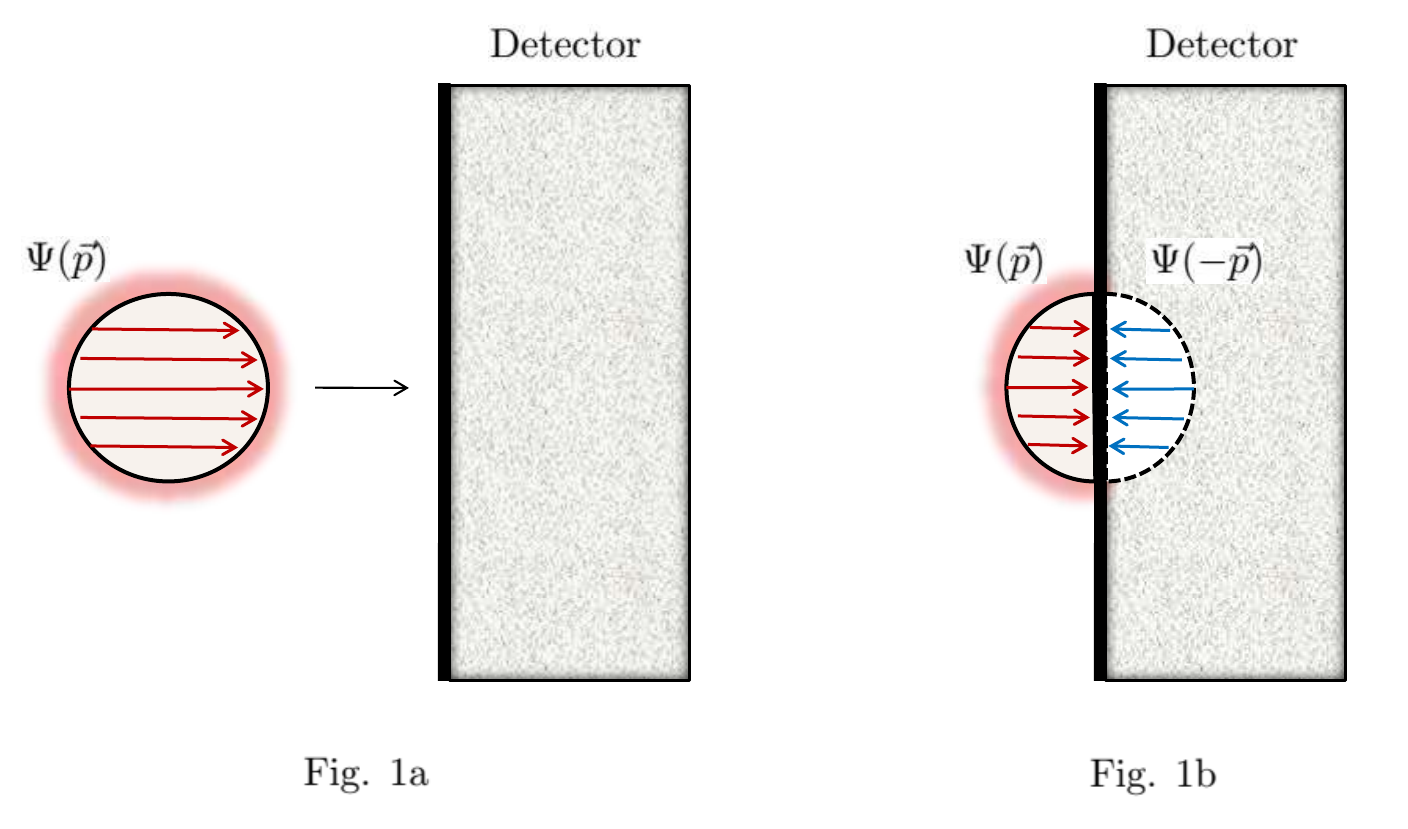}
  \end{center}
  \vspace{-6mm}
  \caption{\small A schematic picture of a free particle and detector (1a) and their interaction (1b).}
%  \center{\includegraphics[width=0.9\linewidth]{fig.pdf}}
%   \vspace{-0.9cm}
%  \caption{\small A schematic picture of a free particle and detector (1a) and their interaction (1b).}
%\label{psi-1}
\end{figure}%\\

The probability amplitude for detecting the particle at some point $x$ with momentum $\vec{p}$ (with the accuracy allowed by the uncertainty principle)
can be written as
\begin{equation}
\label{6}
\Psi_\text{part+ap}(p,x)=\Psi_\text{part}(p,x)\,\Psi_\text{ap}(p,x),
\end{equation}
where $\Psi_\text{ap}$ is the "wave function" of measuring apparatus in the region of interaction (that part of device is often called "meter"), see Fig.~1b.
The key point is the functional dependence of $\Psi_\text{ap}$ on the coordinate and momentum. The full momentum of the particle plus apparatus is $\vec{p}=\vec{p}_\text{part}+\vec{p}_\text{ap}$. Since it is conserved we have
$0=\vec{F}_\text{part}+\vec{F}_\text{ap}$, where $\vec{F}=\frac{d\vec{p}}{dt}$ is the acting force. This is, of course, Newton's third law.

Let us divide the interaction region in Fig.~1b determined by the distribution $f_\text{part}(p,x)$ into many very small segments and apply this law to each segment.
Inside each segment, we will have $\Delta \vec{p}_\text{part}=-\Delta \vec{p}_\text{ap}$ for the corresponding change of momentum. Up to the uncertainty relation,
the point $x$, where this change of momentum happens, can be considered as common for the particle and apparatus.

The consideration of energy change will similarly lead to $\Delta E_\text{part}=-\Delta E_\text{ap}$: If a small segment had the energy $\delta E$ then the energy change
is $\Delta E=0-\delta E$ (as this segment of the wave packet disappears converting to a part of detector), while the corresponding "part of apparatus"
experiences a change in energy $\delta E-0=-\Delta E$. Since the phase is given by $\varphi=Et-\vec{p}\vec{x}$, we
conclude that $\Delta \varphi_\text{part}=-\Delta \varphi_\text{ap}$.

Now these segments can be summed over the whole distribution. The initial phase $\varphi_\text{part}$
is not relevant and can be ignored. This consideration suggests that at the moment of "measurement" one has
\begin{equation}
\label{7}
\Psi_\text{ap}(p,x)=\Psi_\text{part}(-p,x)=f_\text{part}(-p,x)e^{-i\varphi}=\Psi_\text{part}^*(p,x).
\end{equation}
Here it is taken into account that $f_\text{part}(p,x)$ is an even function of momentum $p$ since the underlying Hamiltonian must be quadratic in $p$.

The relation~\eqref{7} is the key proposition. In other words, the particle interacts with the detector in such a way that, thanks to Newton's third law,
its wave function creates its own "image" with opposite phase at the moment of interaction.

This point needs a clarification: The relation~\eqref{6} makes sense only at the moment of interaction of quantum object with a measuring apparatus.
From a practical point of view, the physical meaning of a state vector $\Psi$ is that it represents simply a summary of the past information relevant for future predictions, i.e., a convenient shorthand for all
available information on how the corresponding quantum object (system) was prepared\footnote{The given statement was among the main conclusions in the influential work
by E.~Wigner~\cite{wigner} on quantum measurements. A comprehensive discussion of this standpoint is contained in Peres' monograph~\cite{peres} (summarized at a popular level in~\cite{peres2}), from which it follows, in particular, that assigning any additional objective meaning to $\Psi$ (as suggested by many
interpretations of QM) is beyond the experimental scope.}. For this reason $\Psi_\text{ap}$ in~\eqref{6} has not
existed before the interaction, the latter "prepares" $\Psi_\text{ap}$ following the recipe $\Psi_\text{ap}=\Psi^*_\text{part}$, within the present consideration.

Now a new physical meaning to the relation~\eqref{3} can be given. The relation~\eqref{1} can be interpreted as the {\it necessary} condition for quantum
measurement --- it shows what we should potentially obtain in the case if a measurement is carried out. But the actual measurement is performed only when there was
a "response" from the side of measuring apparatus. To account for this "response", we must multiply~\eqref{1} by $\Psi_\text{ap}$ and integrate over all possible $y$.
What is then obtained is nothing but the relation~\eqref{3} since, according to the reasoning above, $\Psi_\text{ap}=\Psi^*$, where $\Psi$ enters~\eqref{1}.
One can say that the relation~\eqref{3} represents the {\it necessary and sufficient} condition for a quantum measurement of the observable $\lambda$.

\section{Quantum probability}

Admitting that an observable quantity always appears as a convolution of the corresponding quantum state and "induced" quantum state of measuring apparatus we get
a new understanding of the physical reason why the probability in quantum theory is given by the square of the wave function or probability amplitude. The latter
will be below called "quantum probability" which is known to obey all axioms of the classical probability theory\footnote{A brief review of various reflections on the "reality"
of quantum probability is contained in~\cite{youssef}.}. The quantum probability refers to free particles (quantum systems) and for this reason cannot be observed
(as any observation implies an act of interaction). The measurable probability represents the product of two probabilities: (the quantum probability $\Psi$ that we
have a quantum state in some $y$-space)$\times$(the conditional quantum probability $\Psi_\text{ap}$ that the corresponding
measurement of this state occurred), i.e., $P=\Psi\cdot\Psi_\text{ap}=\Psi\cdot\Psi^*$.

One can give the following rough classical analogy. Imagine that we throw a ball into a circle and measure the probability distribution $P_1(x)$ of hitting points $x$ inside
the circle. We are able to do this directly, e.g., thanks to a reflected light. Now imagine that the observer is blind, deaf and does not feel vibrations (this mimics the
situation of an observer in the actual quantum world). In order to record the ball hits, the observer needs now some "measuring apparatus". This apparatus should be of similar
nature as the ball (because quantum particles are detected via interactions with another quantum particles inside a detector), for example, it can be just another ball,
and the observer can detect only the hits of these two balls. For the physical correctness of the analogy, the second ball must move in some unpredictable way inside the circle:
a real quantum particle is not located at a certain point $x$ inside a detector due to the uncertainty principle and thermal fluctuations. Thus, the position of the second ball
has its own probability distribution $P_2(x)$. It is clear that only the product of the probability distributions can be measured: $P=P_1 P_2$.

If the circle is small enough, both $P_1(x)$ and $P_2(x)$ would correspond to a random distribution. But what the "blind" observer would measure in this experiment is the
square of random distribution. This analogy can be generalized to the quantum case. For the same reason that a smooth function can be considered a constant in a sufficiently small
neighborhood of any point, the hit of an incident particle at any point of a sufficiently small detector cell can be considered random, as well as the location of a detector
particle with which the interaction occurred in this small cell. The new feature appearing here is the dependence of quantum probability on the momentum --- Newton's third law dictates
then $\Psi_2(p,x)=\Psi_1(-p,x)$, that yields $P=\Psi_1(p,x) \Psi_1(-p,x)$ for the observable probability distribution.

The fact that the observable probability always represents the square of the quantum probability allows the latter to have complex values. But this bilinear nature of observable
probability does not explain {\it why} the quantum probability is complex. An intuitively clear explanation for this complexity is not known\footnote{A frequent formal explanation
is that the Schr\"{o}dinger equation includes "$i$", $i\hbar\partial_t\Psi=\hat{H}\Psi$, which is indispensable for having the conserved norm $\langle\Psi|\Psi\rangle$ and for various
quantum features. On the other hand, the Schr\"{o}dinger equation represents the non-relativistic approximation to the Klein-Gordon-Fock equation which does not include "$i$".
In reality, the complexity of $\Psi$ must be just postulated (as, e.g., in the path integral formulation of QM).}. What is definitely known is that the complex-valued
quantum probability is more economical in description of multiparticle states (see, e.g., a discussion of this point at a popular level in~\cite{miller}) and allows for a much
broader range of physical phenomena. And as it almost always happens in the quantum world, Nature willingly takes advantage of this opportunity. For instance, any real vector
and scalar in the $SO(3)$ rotation group (or in the $SO(4)$ Lorentz group) can be represented as a tensor product of two complex spinors. The physical properties of spinors
may seem counter-intuitive and at first sight it is not clear at all why these spinors must describe something in the real world. But they do describe fermions (and do this in the most
economical way!) leading to a wide range of physical phenomena which would be impossible for vectors and scalars. The spinors, being the "square roots" of vectors, are subject
to the same formal mathematical operations as with vectors. Likewise, complex quantum probabilities, being the "square roots" of usual probabilities, obey the same formal axioms
as classical real probabilities.

In classical physics, the complex numbers are known to be just a convenient tool for description of wave phenomena. In the quantum world, the situation seems reversed:
It is the fundamentally complex-valued quantum probability of finding a quantum system at some point in configuration or momentum space which gives rise to
wave-like phenomena first discovered in the famous double-slit experiments.

\section{Summary}

Why after calculation of some quantum process, e.g., some Feynman diagram(s), do we square the obtained amplitude? The answer proposed in this note is as follows: Because the final state(s)
of particle(s) must be measured. A Feynman diagram (or any other figure of a quantum process) shows the "preparation" of final state $\Psi_f$ (see the footnote~1) from some initial
state $\Psi_i$. But $\Psi_f$ remains a "thing-in-itself" until the particles in the state $\Psi_f$ reach a detector and are measured. Any measurement is a quantum process subject to
Newton's third law in the momentum space. It is argued that exactly this property explains why the probability $P$ (or probability density) of observation of final particles is given by
the square of the complex amplitude, $P=\Psi\Psi^*$: The observable probability $P$ represents the product of two quantum probabilities --- the first one refers to a "prepared"
particle, while the second (conditional) probability refers to "back reaction" of measuring apparatus signaling that the measurement took place. The expression for the latter (the quantum
probability of detecting a particle that had momentum $p$ with quantum probability $\Psi(p)$) emerges from Newton's third law and is equal to $\Psi(-p)$.
In other words, it is argued that the observable probability is given by $P=\Psi(p)\Psi(-p)$. Since $\Psi(p)$ is complex,
$\Psi(p)=A(p)e^{i\varphi(p)}$, its absolute value does not depend on the sign of $p$ thanks to the Lorentz-invariance, $A(p)=A(p^2)$, and the phase $\varphi(p)$
is linear in $p$, one obtains the equality $\Psi(-p)=\Psi^*(p)$ which guarantees that the measured probability $P$ is always real-valued.

The presented physical argument is somewhat heuristic but it might provide a simple intuition about the origin of Born's rule without delving into perplexed
philosophies of various interpretations of quantum mechanics. It would be interesting to develop a more rigorous framework for the presented physical picture
of entanglement (as a result of interaction) of a quantum system and measuring apparatus that directly leads to the Born rule.


\begin{thebibliography}{99}

\bibitem{chen}
M.-C. Chen {\it et al.},
\textit{Ruling out real-valued standard formalism of quantum theory,} Phys. Rev. Lett. \textbf{128}, 040403 (2022).
\bibitem{li}
Z.-D. Li {\it et al.},
\textit{Testing real quantum theory in an optical quantum network,} Phys. Rev. Lett. \textbf{128}, 040402 (2022).
\bibitem{renou}
M.-O. Renou {\it et al.},
\textit{Quantum theory based on real numbers can be experimentally falsified,} Nature \textbf{600}, 625-629 (2021).
\bibitem{renou2}
M.-O. Renou {\it et al.},
\textit{Quantum Physics Falls Apart without Imaginary Numbers,} Scientific American (2023),
https://www.scientificamerican.com/article/quantum-physics-falls-apart-without-imaginary-numbers
\bibitem{avella}
A. Avella,
\textit{Quantum Mechanics Must Be Complex,} Physics (2022),
https://physics.aps.org/articles/v15/7

\bibitem{born}
M. Born,
\textit{Quantenmechanik der Sto{\ss}vorg\"{a}nge,}
Z. Phys. \textbf{38}, 803-827 (1926).

%Gleason, Andrew M. (1957). "Measures on the closed subspaces of a Hilbert space". Indiana University Mathematics Journal. 6 (4): 885–893.
\bibitem{gleason}
A.M. Gleason,
\textit{Measures on the closed subspaces of a Hilbert space,} J. Math. Mech. \textbf{6}, 885-893 (1957).
\bibitem{hartle}
J.B. Hartle,
\textit{Quantum mechanics of individual systems,} Am. J. Phys. \textbf{36}, 704-712 (1968),
\texttt{arXiv:1907.02953}.
\bibitem{farhi}
E. Farhi, J. Goldstone and S. Gutmann,
\textit{How probability arises in quantum mechanics,} Ann. Phys. (N.Y.) \textbf{192}, 368-382 (1989).
\bibitem{deutsch}
D. Deutsch,
\textit{Quantum theory of probability and decisions,} Proc. Roy. Soc. Lond. \textbf{A455}, 3129-3137 (1999).
\bibitem{saunders}
S. Saunders,
\textit{Derivation of the Born rule from operational assumptions,} Proc. Roy. Soc. Lond.
\textbf{A460}, 1771-1788 (2004).
%W. H. Zurek, "Environment-assisted invariance, entanglement,
%and probabilities in quantum physics," Phys.Rev. Lett. \textbf{90}(12), 120404 (2003).
\bibitem{zurek}
W.H. Zurek, \textit{Probabilities from entanglement, Born's rule $p_k = |\psi_k|^2$ from envariance,} Phys. Rev.
\textbf{A71}, 052105 (2005).

\bibitem{landsman}
N.P. Landsman,
\textit{The Born rule and its interpretation} (2008),
https://www.math.ru.nl/~landsman/Born.pdf
\bibitem{schlosshauer}
M. Schlosshauer and A. Fine,
\textit{On Zurek's derivation of the Born rule,}
Found. Phys. \textbf{35}, 197-213 (2005),
\texttt{arXiv:quant-ph/0312058}.
\bibitem{lawrence}
J.~Lawrence and P.~Goyal,
\textit{The Born rule: Axiom or result?,}
Am. J. Phys. \textbf{92}, 371-374 (2024),
\texttt{arXiv:2302.02086}.

\bibitem{hossenfelder}
S.~Hossenfelder,
\textit{A derivation of Born\textquoteright{}s rule from symmetry,}
Annals Phys. \textbf{425}, 168394 (2021),
\texttt{arXiv:2006.14175}.
\bibitem{gogioso}
S.~Gogioso,
\textit{A very simple derivation of Born\textquoteright{}s rule from symmetry,}
Annals Phys. \textbf{453}, 169295 (2023).

\bibitem{wigner}
E.P. Wigner,
\textit{The Problem of Measurement},
Am. J. Phys. \textbf{31}, 6-15 (1963).
\bibitem{peres}
A. Peres,
\textit{Quantum Theory: Concepts and Methods},
Kluwer Academic Publishers (2002).
\bibitem{peres2}
C.A. Fuchs and A. Peres,
\textit{Quantum theory needs no ‘interpretation’},
Phys. Today \textbf{53}, 70-71 (2000).
%https://pubs.aip.org/physicstoday/article/53/3/70/411209/Quantum-Theory-Needs-No-Interpretation

%\bibitem{feynman}
%R.P. Feynman,
%\textit{The Concept of Probability in Quantum Mechanics},
%Proc. Berkeley Symp. on Math. Statist. and Prob. \textbf{1951}, 533-541 (1951).

\bibitem{youssef}
S. Youssef,
\textit{Physics with exotic probability theory,}
\texttt{arXiv:hep-th/0110253}.

\bibitem{miller}
J.L. Miller,
\textit{Does quantum mechanics need imaginary numbers?}
Phys. Today \textbf{75}, 14–16 (2022).
%https://pubs.aip.org/physicstoday/article/75/3/14/2842709/Does-quantum-mechanics-need-imaginary-numbers-A


\end{thebibliography}
\end{document}